\documentclass[twocolumn,showpacs,preprintnumbers,amsmath,amssymb]{revtex4}
\usepackage{dcolumn}
\usepackage{bm}
\usepackage[dvipdf]{graphicx}
\DeclareGraphicsExtensions{.jpg,.pdf,.mps,.png,.eps,.ps,.EPS}
                           \begin{document}
\def\be{\begin{equation}}
\def\ee{\end{equation}}
\def\bc{\begin{center}} 
\def\ec{\end{center}}
\def\bea{\begin{eqnarray}}
\def\eea{\end{eqnarray}}

\title{Emergence of weight-topology correlations in complex scale-free networks}
\author{Ginestra Bianconi$^{1,2}$}
\affiliation{$^1$The Abdus Salam International Center for Theoretical Physics, Strada Costiera 11, 34014 Trieste, Italy \\
$^2$ INFM, UdR Trieste, via Beirut 2-4, 34014, Trieste,Italy} 
\begin{abstract}
Different  weighted scale-free networks show weights-topology correlations indicated by the non linear scaling of the node strength  with 
its connectivity. In this paper  we show that networks with and without weight-topology correlations can emerge from the same simple growth dynamics of the node connectivities  and of the link weights. 
A weighted fitness network is introduced  in which both nodes and links are assigned intrinsic fitness. This model can show a local dependence of the weight-topology correlations and can undergo a phase  transition to a state in  which the network is dominated by few links which acquire a finite fraction of the total weight of the network. 
\end{abstract}
\maketitle

Recently many complex  systems have been described in terms of the underline network structure responsible for  making  one global entity performing specific functions from many otherwise independent elements \cite{B,D,V}. The scale-free degree distribution is a relevant characteristic of a number of complex networks
which originates, for a large number of real networks, from their growing nature and in the "preferential attachment'' of  new links to  well connected nodes. The minimal model which accounts for these properties is the Barab\'asi-Albert (BA) model \cite{BA}.
Nevertheless  since the early days of research in complex scale-free networks \cite{SY} it has been recognized that the links are not equivalent like in the BA model but they have a different weight $w_{i,j}$ for every link  $(i,j)$.   For example in the Internet \cite{V,Boguna} every connection has a given bandwidth, in the coauthorship network \cite{Airports1}  every couple of coauthors  has a different number of  papers together, in the airport network \cite{Airports1} every connection between two airports has a  different traffic and  in the shareholder network \cite{Guido} the  investments involve  different  volumes of shares.
An important quantity to characterize the relevance of a node in a weighted network is its strength \cite{Airports1} defined as  the sum of all the weights of incoming and outgoing links 
$s_i=\sum_j w_{i,j}$. 
Two classes of weighted networks can be identified looking at  the behavior of  the strength of the node with the connectivity $k_i$, 
\be
s_i\propto \left\{\begin{array}{ll} k_i &\mbox{for class I networks}\nonumber \\
           {k_i}^{\theta} &\mbox{for class  II networks}\end{array}\right.
\ee
with $\theta>1$. 
In real systems an example of class I networks is the coauthorship network\cite{Airports1} while example of class II networks are the airport networks \cite{Airports1}  the shareholder networks \cite{Guido} and the Internet \cite{Boguna}.
The proposed classification discriminate between networks in which the connectivity of the node doesn't affect the  weights of its links (class I networks) and 
networks in which the connectivity strongly influence them (class II networks).
While studying specific systems one could find characteristic feature of them,
a common behavior of  different networks indicate an universality which ask for an explanation.
The present  models of weighted networks in the literature generate  class I networks  \cite{Vespignani,Dorogovtsev,Krapivsky} and  class II networks \cite{Almaas,vespi2} by different mechanisms.
In particular the non linear scaling of strength and connectivity is assumed as an hypothesis in Ref. \cite{Almaas} and it is the signature of some non linear coupling  between weight and strength in Ref. \cite{vespi2}.
In  this paper we want to show that it is possible to explain the emergence of  class I and  class II networks under a  common framework. 
We first propose a very schematic model, the preferential strengthening growing network model. In this model both the weights of the links  and the  connectivities of the nodes grow in time: networks of type II emerge only  when the rate $m'$ at which weights are strengthened  is higher than the rate $m$ at which new links are established.
 Then we propose the  weighted fitness model where a fitness variable  is assigned  to each node and to each link of the network accounting for the intrinsic  capacity of nodes to acquire new links and of links to acquire more weight.
This model generates networks of class I or class II  depending on the value of the ratio $m'/m$.
Moreover in the fitness weighted network  we can distinguish  two phases as a function of the ratio $m'/m$. In the first phase all links have an infinitesimal amount of the total weight of the network while in the second phase few connections grab a finite fraction of all the weight in the network.

The basic idea of both  models is that in many systems the weight of the links is evolving in time while the network grows.
For example in the coauthorship network two scientists in time increase the number of papers written together, in the airport network, while more connections are established between new airports and existing ones, the traffic increases for all the existing connections, in the shareholding network the volume of investment  change with time.
Consequently in our model as the network grows with new nodes joining the network, also the total weight of the network increases and the weight of the existing connections is strengthened. For simplicity we assume that the two processes  appear at the same time scale but with different rate.
We have assumed a "preferential attachment" \cite{BA}  mechanism for the strengthening of the connections of the nodes, i.e. links  will increase their weight in proportion of their actual weight.

{\it The preferential strengthening growing network  -}  Given a starting network we assume that at each time a new node arrives in the network and it attaches $m$ links to the existing nodes of the network establishing connections with a minimal weight $w_0$.
At the same time other $m'$ links increase their weight of an equal amount $w_0$. These links are chosen following a preferential attachment driven by the weight of the links.
In particular  at each time step {\it i)} we add a node and $m$ links following the BA model \cite{BA}; {\it ii)} we choose $m'$ links and increase their weight by $w_0$.
We perform the choice of the  links first by choosing a node with probability proportional to its strength $\Pi_i=s_i/\sum_j s_j$, and then by choosing one  of its links with probability proportional to its weight $\Pi_{i,j}=w_{i,j}/\sum_\ell w_{i,\ell}$. 
The mean field\cite{Note} dynamic equation for  the  weight of a link  is:
\bea
\frac{\partial w_{i,j}}{\partial t}&=& w_0  2 m' \frac{w_{i,j}}{\sum_{\ell,\ell'}w_{\ell,\ell'}}.
\eea
The solution of this equation is
$ w_{i,j}=w_0\left(t/t_{i,j}\right)^{\alpha}  $
where $\alpha=\frac{m'}{m+m'}$, $t_{i,j}=\max(t_i, t_j)$ and $t_i$  is the time at which node $i$ was added to the network.
The dynamics of the connectivities is the simple BA \cite{BA} model with $k_i(t)$ following, in mean field,
$\partial k_i(t)/\partial t=m k_i(t)/\sum_jk_j(t)$
having as solution
$k_i(t)=m\left({t}/{t_i}\right)^{1/2}$.
If we define $p_{i,j}$ the probability that a node $i$ is connected to node $j$ ($p_{i,j}=m/(2 \sqrt{t_i t_j})$, the strength of a node $s_i$  is given by $s_i(t)=\sum_j w_{i,j}(t)p_{i,j}$.
Asymptotically in time we found
\bea
s_i\propto \left\{\begin{array}{lr}k_i &\mbox{if}\ \alpha<1/2 \ \ m'<m  \\ k_i \log(k_i) & \mbox{if}\ \alpha=1/2 \ \ m'=m\\
 k_i^{2 \alpha} &\mbox{if}\ \alpha>1/2 \ \ m'>m\end{array}\right.
\label{st2.eq}
\eea
with the  the weights satisfying
$w_{i,j}(t)\propto (k_{min})^{2\alpha}$ where  $k_{min}=\min(k_i,k_j)$
making of this model when $m'<m$ a class I network and  when $m'>m$ a class II network.
Consequently just with this simple model we found that  class II networks appear spontaneously when the rate of the strengthening of the weights is higher than the rate of addition of new links. 
Moreover it is easy to show \cite{BA} that both the degree and the weight distributions are scale-free with
$P(k)=2tm^2k^{-3}$
and $P(w)=tw_0^{1/\alpha}\frac{1}{\alpha}\frac{1}{w^\delta}$ where $ \delta=1/\alpha+1$.
An important quantity to characterize the inhomogeneities of the weights of the links ending on the same node is $Y^2_i=\sum_j (w_{ij}/s_i)^2$\cite{BioFluxes}. When all the weights give a similar contribution  $Y^2_i\sim1/k_i$ when there is one link with a weight $w_{i,j}\sim s_i$ one finds $Y^2_i\sim O(1).$ 
In our preferential strengthening model we found relevant inhomogeneities also in the class I network for $m'>m/3$. In fact we have   $Y^2(k)\sim 1/k$ for $m'<m/3$ while  $Y^2(k)\sim k^{4\alpha-2}$ for  $m>m'>m/3$. For class II networks $m'\ge m$  we have   $Y^2(k)\sim O(1)$ .
\begin{figure}
\includegraphics[width=85mm, height=45mm]{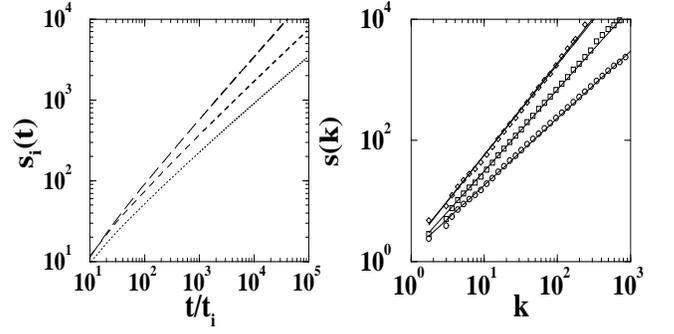}
\caption{Characteristic behavior of the basic model with a  temporal power-law evolution of the strength  $s_i(t)$ (left graph) and power-law dependence of the strength of a node with connectivity (right graph). The data shown are for the preferential streghtening model with in ascending ordering  $(m,m')=(2,1),(2,3),(1,3)$, $N=10^5$ averaged over $100$ realizations.} \label{basic.fig}
\end{figure}
We have performed numerical simulations of the model that are  well captured by the above mean-field treatment. In Fig.$\ref{basic.fig}$ we show the evolution of the strength of a node as a function of time for networks with a different ratio $m'/m$. Moreover we show the non trivial dependence of $s(k)=\langle s_i|k_i=k\rangle$ that deviates from  linearity  if $m'>m$ as predicted by $(\ref{st2.eq})$.

{\it The fitness weighted network-} To justify how correlations between connectivities can emerge  we consider the fitness weighted network.
In this model a random fitness $\eta_i$ from a $\rho(\eta)$ distribution is assigned to each node of the network as in \cite{fitness} and the network grows with new links  attached preferentially to high fitness and high  connectivity nodes.
Moreover  a fitness $\xi_{i,j}$  is assigned to each link $(i,j)$ of the network and as the network grows following \cite{fitness} links with higher fitness and higher weights are preferentially  strengthened.
The distribution from  which is extracted the link fitness $\xi_{i,j}$ will in general depend on the value of the fitness of the two linked nodes $p(\xi_{i,j}|\eta_i,\eta_j)$. 
For this fitness weighted model the weight of single links will increase following, in mean field,
\bea
\frac{\partial w_{i,j}}{\partial t}&=&2m'w_0 \xi_{i,j}\frac{w_{i,j}}{\sum_{\ell,\ell'} \xi_{\ell,\ell'} w_{\ell,\ell'}}.
\eea
Following the same technique used for the fitness model we assume that 
$\sum_{\ell,\ell'} \xi_{\ell,\ell'} w_{\ell,\ell'}=2m'w_0 C't.$
With this ansatz the solution of this equation is straightforward and is given by
\be
w_{i,j}=w_0\left(\frac{t}{t_{i,j}}\right)^{\alpha(\xi_{i,j})}
\label{solwf.eq}
\ee
with $\alpha(\xi_{i,j})=\frac{\xi_{i,j}}{C'}$, $t_{i,j}=\max(t_i,t_j)$ where $t_i$ indicates the time at with node $i$ was added to the network.
Observe that $C'$ has the lower bond $C'<\xi_M=\max{\xi_{i,j}}$ since the weight of a link for construction cannot grow faster than linearly.
For the dynamics of new connections  we use the  fitness model\cite{fitness}. The evolution of the connectivity of a node $i$ is then given by
$\partial k_i(t)/\partial t=m\eta_i k_i(t)/\left(\sum_j\eta_j k_j(t)\right)$
with solution $k_i(t)=m\left(\frac{t}{t_i}\right)^{\eta_i/C}$
and  $C $ given by the self-consistent equation for $C$\cite{fitness}
$1=\int d\eta \rho(\eta) \eta/(C-\eta)$.  Substituting $(\ref{solwf.eq})$ in  $\sum_{\ell,\ell'} \xi_{\ell,\ell'} w_{\ell,\ell'}$ and assuming that this sum is self-averaging, we get the equation for  $C'$
\bea
\frac{m'}{m}&=& \int d\eta_i \rho(\eta_i)\int d\xi_{i,j} p(\xi_{i,j}|\eta_i) \frac{1}{\frac{C}{\eta_i}-1}\frac{1}{\frac{C'}{\xi_{i,j}}-1}.
\label{selfc}
\eea
Note that the right hand side of this expression has clearly its maximum for $C'=\xi_M$ that is the minimum possible value of $C'$.
On the other side the strength  $s_i$ of a node $i$, with fitness $\eta_i$
 can be well approximated
by the value of its average on the fitness of the other nodes and the average on the fitness of its links.
In this approximation we have, asymptotically in time
\be
s_i (k) \propto \left\{\begin{array}{lll} k_i  & \mbox{for}   & \eta_i>\frac{C}{C'}\max_j{\xi_{i,j}} \\  k_i^{\theta(\eta_i)} &  \mbox{for} & \eta_i<\frac{C}{C'}\max_j{\xi_{i,j}} \end{array}\right.
\label{skf.eq}
\ee
with 
$\theta(\eta)>1$ depending on the distributions $\rho(\eta),p(\xi_{i,j}|\eta_i,\eta_j)$.
For every specification of the model one find the weighted scale-free network in class I or in class II depending on the value of $m'/m$.

Let us for a moment consider the case of uncorrelated $\{\eta\}, \{\xi \}$ distributions with $\rho(\eta)$ and $p(\xi)$ uniform with $\eta\in (0,\eta_{M})$ and $\xi\in(0,\xi_M)$. In this case we have that $(\ref{skf.eq})$ holds with $\theta(\eta)=\xi_{M}C/(C'\eta)$.
The average $s(k)$ is then given by $s(k)=\int d\eta p(\eta) s_{\eta}(k) P(k|\eta)$ which gives
\bea
s(k)&\sim&\left\{\begin{array}{lll}   k &\mbox{if}&  C'\ge \xi_M \frac{C}{\eta_M} \\
  k^{\theta}&\mbox{if} & C'< \xi_M \frac{C}{\eta_M}\end{array}\right.
\eea
with $\theta=-C'/{ \xi_M}+C/\eta_M+1>0$.
Since $C'$ is a monotonic function of the ratio $r=m'/m$ this means that there will be a certain value $r_0$ such that for $r<r_0$ the network is in class I and for $r>r_0$ the network is in class II.
As in the preferential strenghtening network model here we found that $Y^2(k)$ indicates relevant heterogeneity also in class I networks. In fact we have for class I networks $Y^2(k)\sim 1/k$ for $C'>2x_MC/\eta_M$ and $Y^2(k)\sim k^{-\lambda}$ with $\lambda\in(0,1)$ and $\lambda=\min(C'/\xi_M+C/\eta_M,-1-C'/(2\xi_M)+C/\eta_M)$ for $C'\in(\xi_MC/\eta_M,2\xi_M/\eta_M)$. For  class II networks instead we found  $Y^2(k)\sim O(1)$.
\begin{figure}
\includegraphics[width=75mm, height=55mm]{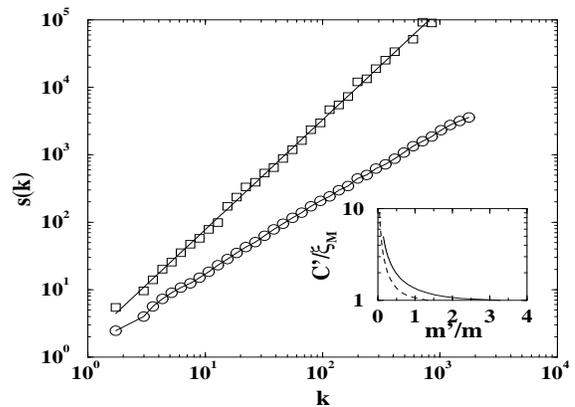}
\caption{Power-law distribution for the strength in the weighted fitness model as a function of $k$ for $\xi_{i,j}=\eta_i+\eta_j$ and uniform distribution of the $\eta\in [0,1]$ for the parameters $(m,m')=(2,1),(1,10)$. The network have  $10^5$ nodes and the results are averaged over $100$ realizations. Inset:  Solution of  $(\ref{selfc})$ as a function of the ratio $m'/m$ for uniform distribution of the $\eta$ and $\xi_{i,j}=\eta_i+\eta_j$ (solid line),$\xi_{i,j}=\eta_i\eta_j$ (dashed line). \label{skC.fig}}
\end{figure}
\begin{figure}
\includegraphics[width=65mm, height=55mm,clip=]{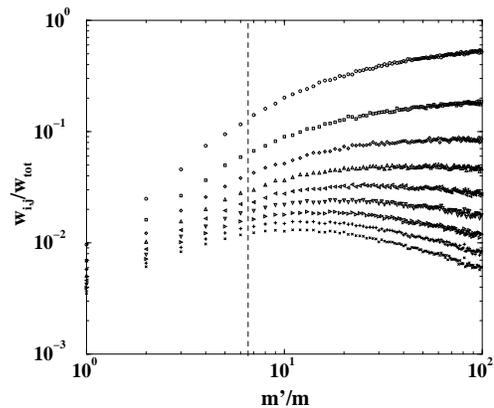}
\caption{Fraction of the total weight covered by the top $9$ links of the network in a weighted fitness model of $m=1$ as a function of $m'$. The data are shown for a graph of $N=10^3$ nodes averaged over $500$ realizations. The dashed line indicates the value $m'/m=3.28$ over which $(\ref{selfc})$ doesn't have a solution and we predict that few nodes grab a finite fraction of the total weight.}\label{cond.fig}
\end{figure} The above treatment is valid as long as the  solutions of  $(\ref{selfc})$ and the self-consistent equation for $C$ exist, i.e.  $\rho(\eta_M)> 0$ and $p(\xi_M)>0$. When on the contrary $\rho(\eta)\sim (\eta_M-\eta)^k$ with $k>0$ we have a phase transition (Bose-Einstein condensation on complex networks -BECCN)\cite{Bose}in which  one node grabs a finite fraction of all the links. Similarly if $p(\xi_M)\sim(\xi_M-\xi)^{k'}$ with $k'>0$ we have a  ``Bose-Einstein condensation'' of the weights of the network, i.e. one link grab a finite fraction of all the weight of the network. The study of a simple  case of unlimited support of the probability distributions $\rho(\eta)=e^{-\eta}$ (or $p(\xi)=e^{-\xi}$) can be easily be carried out \cite{fitness}.Here we observe that if both fitness of the nodes and fitness of the links are exponentially distributed, the scaling  $(\ref{skf.eq})$ remains valid. 

Even more interesting is the case in which  the fitness of the links  is functionally related to the fitness of the nodes $\xi_{i,j}=f(\eta_i,\eta_j).$ This  is a natural assumption in many real complex systems. For example, more productive scientists will produce more papers both with existing coauthors and  with new collaborations, airports which attract more connections will reasonably  also be the airports which attract more traffic, richer agents will invest more in new investments and in their present portfolio.    
Two relevant examples of shape of $f(\eta_i,\eta_j)$ would be for example
$f(\eta_i,\eta_j)=\eta_i+\eta_j$ and $f(\eta_i,\eta_j)=\eta_i \eta_j$.
We will first   consider the case of  additive   $f(\eta_i,\eta_j)=\eta_i+\eta_j$ with uniform distribution of the $\eta$ and $\eta\in(0,\eta_M)$. In this case  $C/\eta_M=1.255\dots$ \cite{fitness} and $C'\leq 2\eta_M=x_M$. In Fig. $\ref{skC.fig}$(Inset) we plot the value of $C'$ as a function of the ratio $m'/m$ as predicted by  $(\ref{selfc})$ in this case.
In $C'=2\eta_M$ the integral in $(\ref{selfc})$  converge allowing
for the maximal value of the ratio $\frac{m'}{m}=r_c=3.28\dots$ for  $(\ref{selfc})$ to be valid.
This scenario is indeed reminiscent of BECCN.
In fact  the convergence of $(\ref{selfc})$ for $C'=2\eta_M$ at $m'/m=r_c$ is the signature that for higher values of the ratio $m'/m$ there will be
 some links that have a finite fraction of the overall weight of the network.
We performed numerical simulations of the fitness weighted network model as a function of the ratio $m'/m$. In Fig. $\ref{cond.fig}$ we plot the fraction   over the total weight of the weight of the top links of  networks fo size $N=10^3$. We observe that for $m'/m>r_c=3.28\dots$ the fraction of weight acquired by few top nodes is a finite fraction of the total weight.
The difference between this transition and  BECCN is that in the BECCN a single node  grabs a finite fraction of the links while  here we have few links that grab a finite fraction of the total weight of the network. In fact we have a finite density of links with maximal fitness $p(\xi_{i,j}=2\eta_M)>0$. Below this phase transition  we found  $s(k)\sim k^{\theta}$ with $\theta=1+(2 C-C')/\eta_M$ for $C'>2C$,i.e. the network is a class II network for $m'/m>r_0=1.18\dots$ and $s(k)\sim k$  for $C'<2C$, i.e. the network is a class I network for  $m'/m<r_0$. 
In Fig. $\ref{skC.fig}$ we report the average strength of node as a function of the connectivity $k$, $s(k)$ for network in class I ($m'/m=0.5$) and in class II $(m'/m=10)$.
The degree  and the weight distribution below the phase transition
follows a power-law this time with a logarithmic correction
$P(k)\propto\frac{k^{-\gamma}}{\log(k)}$ and $P(w)\propto\frac{w^{-\delta}}{(\log(w))^2}$
with $\gamma=1+C=2.255\dots$ and $\delta=1+C'/2$ while above the phase transition in the degree and weight distribution is show the presence of few condensed links.

This scenario we are presenting  is not restricted to the specific case  $\xi_{i,j}=\eta_i+\eta_j$. For example the phase transition indicating the ``condensation'' of the weights is    also present in the case of the multiplicative dependence of the fitness of the links with the fitness of the nodes $\xi_{i,j}=\eta_i\eta_j$ and uniform $\rho(\eta)$ distribution with the critical ratio  $r_c\sim 1.334$. In the Inset of Fig. $\ref{skC.fig}$ we report the dependence of $C'$ on the ratio $m'/m$ for this last case. In this case the network is a class I network for $C'>C\eta_M$ and a class II network for $C'\leq C\eta_M$ with $\theta=C\eta_M/C'$ and the critical value of $m'/m=r_0=0.59$.

In conclusion we presented a general framework for weighted networks which describe the emergence of weight-topology correlations. 
In this framework  both  node connections  and link weights  increase following a preferential attachment rule.
The networks of  class I  or class II (with significant weight-topology correlations) are obtained as a function ratio $m'/m$ between the rate $m'$ at which links are strengthened and the rate $m$ at which new connections are established.
Moreover also in class I networks there can be inhomogeneities in the weights of node edges  reflected in a  non trivial $Y^2(k)$.
For the weighted fitness model which accounts for networks with degree-degree correlations the hetereogenieties in the weights of the nodes can be so strong that for $m'/m>r_c$ there is a structural phase transition to a phase in which few links grab a finite fraction of the total weight of the network.

We acknowledge M. Barth\'elemy, G. Caldarelli, M. Marsili and A. Vespignani for stimulating and interesting discussions.


\begin{thebibliography}{}
\bibitem{B}
R. Albert and A.-L. Barab\'asi, Rev. Mod. Phys. 74, 47 (2002).
\bibitem{D}
S. N. Dorogovtsev and J. F. F. Mendes, {\it Evolution of Networks} (Oxford University Press, Oxford, 2003).
\bibitem{V}
R. Pastor-Satorras and A. Vespignani, {\it Evolution and Structure of the Internet }(Cambridge University Press, Cambridge, 2004).
\bibitem{BA}
A.-L. Barab\'asi and R. Albert, {\it Science } 286,509 (1999).
\bibitem{SY}
S. H. Yook, H.  Jeong, A.-L. Barab\'asi and Y. Tu,
    Phys. Rev. Lett. 86,  5835 (2001).
\bibitem{Boguna}
M. A. Serrano, M. Bogu\~n\'a and A. D\'iaz-Guilera, Phys. Rev. Lett. {\bf 94} 038701 (2005).
\bibitem{Airports1}
A. Barrat,M. Barth\'elemy, R. Pastor-Satorras and A. Vespignani, PNAS 101, 3747 (2004).
\bibitem{Guido}
 D. Garlaschelli, S. Battiston, M. Castri, V. Servedio and  G. Caldarelli, cond-mat/0310503.
\bibitem{Vespignani}
A. Barrat,M.  Barth\'elemy and A. Vespignani, Phys.  Rev. Lett. 92, 228701 (2004).
\bibitem{Dorogovtsev}
S. N. Dorogovtsev, and J. F. F.  Mendes, cond-mat/0408343.
\bibitem{Krapivsky}
T. Antal and P.L. Krapivsky cond-mat/0408285 .
\bibitem{Almaas}
E. Almaas, P. L. Krapivsky and  S. Redner cond-mat/0408295.
\bibitem{vespi2}
A. Barrat, M. Barth\'elemy and A. Vespignani, cond-mat/0406238.
\bibitem{Note}
Note that in the following we will always perform  mean field calculation. By mean field calculation   we intend that
the stochastic variables $k_i(t)$ and $w_{i,j}(t)$ are identified with their average value. This approximation can be proven to be work very well for growing networks as the one we are considering in this paper\cite{D}.
\bibitem{BioFluxes}
E. Almaas, B.  Kovacs,T. Vicsek and A.-L. Barab\'asi, Nature 427, 839 (2004).
\bibitem{fitness}
G. Bianconi and  A.-L. Barab\'asi,  Europhys. Lett. 54,  436 (2001).
\bibitem{Bose}
G. Bianconi and A.-L. Barab\'asi, Phys. Rev. Lett. 86, 5632  (2001).
\end{thebibliography}
\end{document}